# How to be a Good 'Frand' in China: an Exploratory Study of Effective Social Media Behaviours for Foreign Brands

**Mary Tate**
School of Information Management
Victoria University of Wellington, New Zealand
Mary.tate@vuw.ac.nz

**Hongzhi Gao**
School of Marketing and International Business
Victoria University of Wellington, New Zealand
Hongzhi.gao@vuw.ac.nz

**Hongxia Zhang**
School of Marketing
Peking University, PRC
hxzhang@gsm.pku.edu.cn

**David Johnstone**
School of Information Management
Victoria University of Wellington, New Zealand
David.johnstone@vuw.ac.nz

## Abstract

Many foreign companies see social media as a low cost marketing space for entering the vast Chinese market. This is fraught with complexity, requiring an understanding of social media behaviour, social media marketing, and the Chinese context. In this exploratory pilot study, we develop the notion of 'franding' (online friending of brands) as a unifying metaphor, and suggest that effective social media brand strategies resemble effective online friend behaviours. Based on this we develop some propositions and evaluate them qualitatively in interviews with young Chinese consumers. Our study suggests that conceptualizing brand social media relationships as a type of online friendship can assist in harmonizing disparate literature and provide useful insights for organizations operating in this complex area.

## 1. Introduction

China is emerging as one of the biggest export markets for Australia and New Zealand, and many companies are seeking ways to enter the Chinese market. Social media is sometimes seen as a promising marketing channel with low entry costs, especially for smaller brands.

A nascent area of study of 'franding' studying the trends and behaviours associated with friending brands on social media is emerging. In this paper, we report a pilot study from a larger program of work on branding in social media in a Chinese context. We use online friendship as a unifying metaphor in this study; we ask how can a brand be a good online friend? In particular, we draw on research in online friendships to frame the study. We know that social media environment offers opportunities to experiment with self-representation and identity construction (Gao, Ballantyne, & Knight, 2010; Kietzmann, Silvestre, McCarthy, & Pitt, 2012), and therefore we explore the possibility the social media context may offer opportunities for experimentation for *brand* personalities as well. We develop the idea that the more a brand entity on a social media site behaves like other social media 'friends' the more likely it will attract attention, acceptance engagement, and positive e-word of mouth.

One of the challenges of conducting research on the use of social media by brands, especially across cultures, is the wide variety of influences (and corresponding theories to explain them) that are involved in the phenomenon. Reductionist approaches using a single theoretical lens may fail to capture and explain the complexity involved. We start from the perspective that people, including public social expression of their networks, taste, personality, peer group, life-stages, entertainment, goals and motivations are the 'heart and soul' of social network systems (SNSs). Although marketers might wish otherwise, engagement with brands is not a major motivation for social media use for most people. However, to the extent that consumption and brands form a valuable addition to a person's digital social life, they can be welcome participants in social network spaces.

We take an integrative approach, drawing on multiple, intersecting theoretical lenses to explain different aspects of online friendship behaviour. In the rest of this paper we offer a literature review of the properties of social media, personality expression on social media, motivations for using social media, and friending behaviour on social media. We apply each of these to the Chinese context, and then develop these into propositions about desirable brand behaviour on Chinese social media. Following this we describe the methodology and our pilot study, and present some preliminary results. Next we offer a discussion of the emerging findings and some conclusions.

## 2. Literature review

We briefly review some of the theoretical lenses that can be used to contribute to understanding effective online friendship and how that applies to brands. We also discuss people's motivations for using social media and engaging with brands online. Next we introduce some of the special features of the Chinese context. Finally we develop some research propositions based on this literature review.

### 2.1. Properties of Social Media

At the broadest level, social media, and any form of digital marketing involves the production and management of *digital* resources. Compared to other business resources, digital resources have specific material properties: they can be addressed and communicated with, and be aware and responsive to their context (for example, many websites recognize who is visiting and make targeted recommendations); they are persistent, replicable and can be traced (for example, a digital conversation between a company and a customer is not confidential but can be readily shared with other customers (Yoo, Lyytinen, Boland, & Berente, 2010)); and all digital objects have the ability to be associated and combined with other digital objects so long as you stay in a digital environment (Yoo, Henfridsson, & Lyytinen, 2010).

SNSs are a special case of digital resources that include all the other properties of digital resources, plus some additional ones. One essential characteristic of SNSs – the visibility of a user's social networks – was identified by researcher and commentator Danah Boyd: Typically social networks allow people to (1) construct a public or semi-public profile within a system; (2) identify a list of other users with whom they share a connection, and (3) view and traverse their list of connections and those made by others within the system (Boyd & Ellison, 2007). In particular, *"What makes social network sites unique is not that they allow individuals to meet strangers, but rather that they enable users to articulate and make visible their social networks"* (Boyd & Ellison, 2007, p. 211). Another important characteristic is mass participation, and increased 'trust' by contributors, representing an understanding that due to the sharable and recombinable properties of digital content, content that was posted might be reused and repurposed by other users.

SNS properties (affordances) create opportunities for brands. The possibilities of interaction between a person and an object, based on the material properties and inherent characteristics of each (for example, the sensory acuity of humans falls within a defined range, and there are some sounds they cannot hear). The material properties of social media software therefore provide a range of 'social affordances' for users – properties and characteristics that support communication and interaction in a multi-media environment (Wang, Meng, & Dong, 2012). This provides the ability for brands to offer rich interaction with users, and entertainment in the form of games and competitions. A study in the Chinese context showed that brands that provided more interactive features, and more social game features attracted more 'frands' (online friends of brand pages) than brands that offered fewer of these features (Wang et al., 2012).

## 2.2. Friending Behaviour

Social capital can be considered as the collective value of all connections among and within social networks. Fukuyama (1995) explained that social capital shows the value of promoting social cooperation and is instantiated in actual social relationships. Participation in SNS may increase one's social capital (Ellison, Heino, & Gibbs, 2006). Conceptualizations of social capital involve patterns of behaviour, information exchange, trust, and reciprocity within social networks (Putnam, 1995).

The existence of SNS can influence the way members maintain social ties and form new connections. Parks (2009) and Ellison et al. (2007), both mentioned two types of social capital in their social network analysis studies, bonding social capital and bridging social capital, and applied these concepts to SNS. Bonding social capital is a strong social tie involving intimacy between individuals, which provides meaning and social support that sustains and promotes individuals (Parks, 2009). Putnam (2000) referred to bonding social capital as a 'tightly-knit' set of relations between individuals. Bonding social capital can be found between individuals who possess emotionally close relationships, such as family and close friends. In a social media context, this usually means that strong ties will be associated with those with whom people have close, existing off-line relationships as well, although there are exceptions. Bridging social capital relates to social ties between individuals with weaker ties due to the diverse individuals and groups within the community (Parks, 2009). As Putnam (2000) saw it, bridging social capital refers to the loose connections between individuals who may provide useful information or new perspectives for one another, but typically not emotional support. Social sites on the internet are theorized to increase bridging social capital by facilitating the development of weak and diverse ties, since friends can be added with a 'click'. SNS may facilitate bridging social capital because weak ties can easily be maintained and activated (Donath & Boyd, 2004; Ellison et al., 2007). Haythornthwaite (2005) discussed latent ties, which are social links that are technically possible but not usually activated socially. SNSs also support latent ties, because it makes one's connections visible to a wide range of individuals, and enables network members to identify those who might be useful in some capacity (and therefore activated into a weak tie).

## 2.3. Personality Expression on Social Media

An individual's construction of their social digital identity is a subset of their ongoing process of identity construction and performance. A basic premise of identity theory is that modern society is a complex and multi-faceted mosaic of interdependent but highly differentiated parts (Simon, 2004). The social person is equally complex. People have multiple identities which result from participation in multiple sets of role relationships. Identity theory addresses both personal identity, or self-concept, and social identity. Self-concept is the generalised self-descriptions and images we have of ourselves, that we use to recognize and interpret stimuli in the social environment (Marsh & Hattie, 1996). Personal identity involves the process of self-definition as a *unique individual* in respect of interpersonal or intragroup definitions (Turner, Hogg, Oakes, Reicher, & Wetherell, 1987). Social identity, by contrast, relates to self-definition as a *group member* in terms of in-group/out-group definitions. In particular, self-categorization theory suggests that social identity results from self-categorizations, which are "cognitive groupings of oneself and some class of stimulus as the same" (Turner et al., 1987, p. 44), and distinct from other classes of stimulus. People may be members of many different groups, but not all groups are equally salient to the person at any time.

Social identity in the off-line world is recognised by behaviour and physical representation. In computer-mediated communication (CMC), the performance of identity occurs primarily not through direct experience of the body but within the constraints of digital representations constructed by interactive systems. Digital social identity is more explicitly articulated by the owner, who has a greater degree of control over what they choose to present in a digital context (Boyd, 2003). Digital identity does not have to be 'true' but must be credible to peers (Boyd, 2003). Social media identity can be aspirational and express an idealised self. As we discussed earlier, while self-categorization into groups is very easy, even instantaneous, it may be very superficial and involve a large number of weak ties. Nevertheless it is also very persistent, past choices remain and can be viewed and searched. The user's choice to disclose their group memberships (friendship groups, of brand communities) is therefore part of identity performance.

## 2.4. Brands and Personality Expression in Social Media

It has been suggested that the range of possible identities a person can construct is bounded by models, images and symbols provided by the media, in conjunction with the individual's immediate social experiences. From these possibilities, the individual will self-categorize according to the perceived salience of the concept presented to them (Bennet, 2012). Brand identities form part of the range of images and symbols that are available to the individual to use in their identity construction. Product and brand choices can be used both to construct our self-concept (self-symbolism) and our social identity (social symbolism) (Elliot & Wattanasuwan, 1998). For example, selecting an international brand may contribute to our self-concept as being sophisticated and attentive to quality, or may reinforce memories of overseas travel. Through the socialization process, consumers learn not only to agree on shared meanings of some symbols, but also to develop symbolic interpretations of their own. Consumers use these symbolic meanings to construct, maintain, and express each of their identities. The notion of 'discursive elaboration' from advertising theory is highly relevant to social media communities. If we apply the corresponding aspect of SNSs (in square brackets) to advertising theory (O'Donohoe, 1994), we see that the process of discursive elaboration involves the social consumption of advertising [on social media] meanings. [Brands in social media spaces] become tokens in young people's system of social exchange; they are a form of cultural capital and can be invested to gain social status and self-esteem.

This is supported in a study by Wallace et al. (2012), who proposed that homophily (the principle that contact between similar people occurs at a higher rate than among dissimilar people) contributes to constructing strong social ties. Wallace et al. (2012) found that users with strong social ties on SNSs were more likely to use brands for symbolic self-expression, and to engage in brand advocacy.

## 2.5. Motivations for Using Social Media

Users have a wide variety of motivations for using social media. As well as self-expression, studies based on use and gratification theory have found that users seek social communication with friends, social support, information, entertainment, and convenience (Kim, Sohn, & Choi, 2011). An interesting industry-oriented study (Baird & Parasnis, 2011) investigated user motivations for interacting with companies on SNSs, and found that these included (in order from most to least important): purchasing goods, obtaining discounts, obtaining various types of information (corresponding to the information motivation above), submitting opinions (social communication), obtaining service (convenience), while motivations such as being part of a community (social support) were in last place. Further, Baird and Parasnis (2011) found that while the motivations that companies assumed users had for engaging with them on social media were similar, companies mistakenly assumed that users' priorities were almost the reverse of what they actually are.

## 2.6. The Chinese Context

The properties and nature of digital resources are arguably not markedly different in the Chinese context. Chinese SNSs are a largely parallel universe, with leading western social media sites such as Facebook banned in China, while the Chinese market is completely dominated by home-grown offerings such as WeChat, RenRen and Weibo, which fill similar niches to products such as Facebook and Youtube. China has the largest overall numbers, and some of the most active social media users in the world[1].

Friending behaviour is based on social networks in China, as it is in Western contexts, but conceptualization of social ties has a specifically Chinese flavour. Social and business relationships in China are understood in terms of 'guanxi' which translates to 'relationship' in Chinese. However, specifically 'guan' in Chinese means 'gate'. People inside the guan are in the in-group, while people outside the guan are excluded. This mirrors the in-group/out-group concept from Western social psychology. Guanxi can be seen as having three levels of ties (Hwang, 1997). The innermost circle consists of people connected by kinship and closest friends. The next circle includes friends and people who are considered close, while the outer circle includes people who just know each other. The outermost circle is similar to the concept of weak ties from social network theory, in that those people are considered to have a mainly instrumental, rather than emotional connection (Gao et al., 2010).

In a business relationship network context, some further cultural differences exist that are relevant to this study. Chinese businesses can be more short term, instrumental and 'deal based' than Western business relationships, especially when no close guanxi relationship exists, as they may be less interested than western organizations in creating long term relationships and partnerships (Gao et al., 2010). This may suggest that short term brand offers such a discounts and deals are more likely to be received favourably in a Chinese SNS context than membership of brand communities.

In terms of online behaviour, a study by Guan and Tate (2013) found that even in western professional contexts, the introduction of 'boundary crossing' relationships on SNSs (for example, a lecturer 'friending' students) caused changes in behaviour and reduced willingness to disclose personal information and feelings. The culture of high power distance that exists in China suggests that online social networks may place more emphasis on homophily (close connections between similar types of people), and be even less likely to include 'boundary crossing' relationships. This suggests that the online preferences and behaviours of the peer group are extremely likely to influence SNS users in a Chinese context.

In terms of self-expression, we expect that expressed social media personalities will reflect the diversity of modern Chinese culture. In the large coastal cities, many people have an international outlook. They may have travelled overseas, or aspire to. Food franchises such as KFC and McDonalds are well-established in China. For Chinese youth, these chains represent an opportunity to participate in a global youth culture, and an opportunity to meet and socialize with friends (Eckhardt & Houston, 2001). The purchase of luxury brands can be used to communicate social status and earn respect (Zhan & He, 2012). It is possible that conspicuous consumption and identification with luxury brands may also be evident in social media profiles.

Interestingly, the ability to experiment with different online personalities is available to brand pages as well as individuals on SNSs. Foreign brands may present themselves as aligned with local and national culture and traditions. This is known as 'glocalization' where global brand values are incorporated into local consumption practices. This can assist in making non-traditional products such as coffee, lamb, or yoghurt more appealing to Chinese consumers. In addition, brand personalities can be developed, extended and experimented with on SNSs. For example, the ability to post videos offers the ability for brands normally considered as commodities to develop a 'personality'.

---

[1] http://www.statista.com/statistics/278414/number-of-worldwide-social-network-users/

## 2.7. Research Propositions: Foreign Brands as Effective Frands in China

Based on these insights from research, we developed a number of theoretical propositions showing how effective social media strategies for foreign brands in China can be based on understandings of how to be a good social media frand.

*P1: Effective frands will include compelling digital content, and encourage sharing and recombining brand content with personal content.* The sharable, addressable, recombinable attributes of digital resources (Yoo, Henfridsson, et al., 2010) will be utilized both to develop compelling digital content, and the social affordances will encourage users to provide content to the brand and repost brand content in their own profiles (Boyd & Ellison, 2007).

*P2: Effective frands will reflect appropriate social network and guanxi tie strength.* The closest circle of guanxi relationships online will be associated with close offline relationships with kinship groups and closest friends (Gao et al., 2010), echoing the concepts of 'close ties' (Parks, 2009) and 'bonding social capital' (Putnam, 2000) from social network theory. Only the most loyal customers have an emotionally close relationship with a brand, in the best case it may potentially exist in the second circle of guanxi relationships offline. While close engagement may be a goal for brand managers, the majority of their customers will have weak or latent ties with the brand, in other words, the brand will be in the outermost guanxi circle (Gao et al., 2010). In this circle, guanxi connections have a more instrumental, less emotional focus. We suggest that effective online friending behaviour by brands should reflect social and guanxi tie strength, and not demand more intimacy from customers than is appropriate. This implies that brands may be better to offer functional and instrumental benefits on SNSs to reflect the motivations of the majority of customers with whom they are weakly connected.

*P3: Effective frands will reflect the online preferences and behaviours of the customer's peer group.* Effective brand presences on SNSs demonstrate understanding of the customer by the interests, values and concerns of the target customer's peer group. They are associated with events, other brands, and symbols that are meaningful in the customer's network. A culture of power distance is likely to reduce boundary crossing social media relationships, and demonstrate 'homophily' within the peer group (Wallace et al., 2012), more important for Chinese consumers. In terms of brand identification, cultural symbols and consumption behaviour, the social identity of the peer group is expected to be important.

*P4: Effective frands will encourage personality expression and experimentation by both the brand and the customer.* SNSs allow the construction of personality and identity, and the user's self-categorization into friendship groups, or brand communities is a form of social self-expression. However, since online personality expression is considered to be a construction and performance by the user, to a greater degree than offline personality expression, it also encourages experimentation. It is also easy and almost instantaneous to form weak connections by 'liking' a page. Online symbols offered by the brand, expressed in digital materials such as photographs, music and videos, as well as text posts, can be incorporated into the user's online self-expression.

Both the user, and also the brand, can express their personalities by the symbols they incorporate, and the social networks they create. For example, the brand may choose to be associated with celebrities or events as part of its online personality expression. The brand may also 'experiment' with the incorporation of Chinese symbols into its Chinese SNS personality to build a glocalization strategy.

*P5: Effective frands will reflect and support users' motivations and priorities for using SNSs.* Friends help each other and are not narcissistic and self-absorbed. Therefore the brand SNS presence should not concentrate on 'talking about itself' but on the motivations and priorities of its customers for using a SNS. Typically, brand intimacy and emotional engagement with a brand community are not very high on the customer's list of priorities for engaging with a brand on social media (Baird & Parasnis, 2011). Instead, users are more likely to be interested in purchasing goods, obtaining discounts, obtaining various types of information (corresponding to the information motivation above), submitting opinions (social communication), and obtaining service (convenience). A brand's SNS pages should support users to achieve their goals. The brand SNS page can also be a good social network friend by supporting users' wider goals for SNS use (Kim et al., 2011), and offering general information of interest, entertaining content, and assistance to make daily life more convenient.

## 3. Methodology

To gain some empirical insights we conducted an initial exploratory study consisting of in-depth interviews with twelve young consumers in Beijing in 2013. We used 'snowball' sampling, starting with students at Renmin University. Seven interviewees were female and five were male. All were aged between 18 and 30. All were currently resident in Beijing, but the majority (nine) had moved to Beijing from other regions to attend university, and were from diverse family backgrounds, including farming, public service, and business. Semi-structured interviews were conducted on site. Interviews were conducted primarily in English, but Chinese was also used to explicate key points. The interviews were open and exploratory, and invited our participants to share their experiences of brands that they were aware of, or engaged with, in social media. We used our participant's willingness to 'friend' a brand as a proxy for the effective SNS presence and behaviour of the brand.

Ensuring a representative sample is challenging in a country as large and diverse as China. Marketers have suggested that China can be characterised into nine major segments or 'nations' (for example, the north-east 'rust belt', or the south-west 'Shangri-La'), as even when considered individually, each segment would be in the top twenty countries in the world by population (Chovanec, 2013). While the availability, and appetite for imported goods is growing rapidly throughout all regions of China (Economist, 2014), foreign brands still tend to have the largest presence in the large coastal cities (Economist, 2014), and are most popular with affluent urbanites. In addition, we needed to recruit interviewees who were fluent in English, engaged with foreign brands and social media culture, had a strong peer group, yet represented a diversity of backgrounds. Therefore, while we make no claims that our sample can represent the entire Chinese market, it is relatively representative of people with the experience and characteristics that were of interest for our study.

In the first stage, the semi-structured interviews were transcribed, Chinese language sections were translated, and the transcripts read by the research team. Online brand communities were also visited. In the second stage, codes were generated across the whole dataset, based on categories generated from literature, collating data relevant to each code. In the third stage, the codes were organized according to the research propositions. In the fourth stage, the codes and propositions were reviewed. Some extracts were recoded and some codes were amalgamated. In the fifth stage, the result set and emerging stories were refined to ensure that the overall story of the analysis was truthful and consistent. Finally, selected extracts were chosen for inclusion in the report. Multiple informants, and direct observation of some of the brand communities discussed allowed us to triangulate data, to authenticate accounts, and also to understand the different perspectives, experiences and meanings of the interviewees.

## Results

*P1: Effective frands will include compelling digital content, and encourage sharing and recombining brand content with personal content.* The multi-media capabilities of social networks were remarked on; frands offering entertaining video content were popular, R3: *"[Some brands]…will shoot micro movies…They were trying to keep your eye on it and make you desire to continue watching it….they also invite few famous stars to attract people watching it."* Similarly, the recombinability of social media was noted as part of 'franding' behaviour by restaurant SNSs, R4: *"if some customers have dinner there, they will post some photos and the company will repost them."* The interactive possibilities and social affordances of SNSs, including the opportunities for self-expression and co-creation combining brand and personal content, were also remarked on. R6 commented on Nike's shoe design competition: *"If I design shoes and I think it was really cool, I just put the picture up….I like to show people*

that I like red colour or I like purple colour..." or R10: "*[I follow brands that] have interesting activities. Like design or competitions.*"  R11 summarized it well "*If the company holds an interesting activity or shoots interesting short movies, I will tell my friends on social media*". Overall, unsurprisingly, this proposition was supported. Leveraging the properties and social affordances of social media, effectively with interesting content (such as videos) or interesting interactive activities that encourage participation, will encourage customers to frand brand pages.

*P2: Effective frands will reflect appropriate social network and guanxi tie strength.* An interesting insight from this research is that despite the known social symbolic dimensions of consumption, we are not necessarily deeply engaged with many of the brands we consume and interact with. Many of our respondents reported 'weak tie' or 'outer circle' guanxi relationships with online brands.  R4 reported a mainly instrumental relationship with online brands: "*There is only small possibility that I will follow [a brand on social media]. Because so far I couldn't see any brand that really draws my attention. For example, in Wechat, if they send you some information, and ask you to respond. They will reward you with a gift or go for a prize draw, which means it's worthwhile to follow their brand. So if they do promotions, which draws our attention or our interests, then we will follow them.*" However even free offers need to be relevant to be appealing, R8: "*There are also many brands doing it, maybe just not attractive to me. The ordinary ones, you can @ your friend to get some companies' products, but the product may not be the ones that I want, such as a toilet cleanser.*"

However deep engagement with online brands as a form of symbolic self-expression was also reported, although it was rarer. R7 described a friend who was deeply engaged with the Chanel brand. The consistency and authenticity of the relationship was important to her friends on social media: "*I know a friend whose favourite is Chanel. So maybe she has the same personality with the brand. If this person always post the pictures of wearing some Chanel necklace or bag or shoes, that means she loves Chanel. She is connected with Chanel.*" It is recognized as a strong tie relationship, not merely as conspicuous consumption: "*But if a person just post different brands, one day is Chanel, one day is Gucci and the other day is Amani. It just means the person likes to follow trends and show off. This person, not only she is rich, but there is something in her heart.*" Another respondent identified closely with IBM, and was deeply engaged with their Weibo presence because it expressed a vision that resonated with his values: "*I think IBM is trying to be objective towards the society and it has its own vision. They use social media such as Weibo to communicate their ideas, especially they share the 'smart planet' which is a vision about future life to make society functional and regulations easier, made the environment cleaner and lots of other things…I believe in that vision.*"

In another interesting parallel with social ties between people, many of our respondents reported developing an offline relationship with a brand first, before considering franding it. R5, who followed Pepsi online, was already a fan before starting an online relationship, and described the evolution of the relationship: "*Pepsi uses many top football player to endorse their product. I saw many ads about Pepsi. I attended many activities sponsored by Pepsi. Gradually I became a fan of Pepsi.*" Many respondents, although we were discussing social media relationships, mentioned packaging, signage, and advertising through other channels as major sources of brand awareness that preceded online engagement.

Overall, this proposition was supported. Many brand campaigns effectively leverage weak ties and instrumental, outer guanxi circle relationships. Many customers only engage superficially with brands. A much smaller group of people will have a stronger, more emotionally involved relationship that forms part of their digital identity and self-expression. These relationships need to be very authentic or they run the risk of being judged as superficial and materialistic by their friendship group. Also, we found that with very few exceptions, online frand relationships were preceded by offline relationships and awareness of the brand, and consumers were very cautious about investing their online identity in a brand. This may not be very good news for smaller companies hoping to use social media as a major marketing channel, but it appears that it is hard to make 'online only' friends. Nevertheless, supporting previous research on social media and latent ties, the reach provided by social networks is very large. The extent to which these latent ties to 'friends of friends' are valuable to organizations requires further investigation.

*P3: Effective frands will understand and reflect the online preferences, behaviours and issues of the customer's peer group.* Effective brand presences on SNSs demonstrate understanding of the customer by the interests, values. Our interviewees were university students from a wide range of socio-economic backgrounds. While conspicuous consumption might be popular in some sectors of Chinese society, it was frowned on by our interviewees. R11: *"I don't do that [post pictures of purchases]. Because it seems to people that I am too materialistic...It depends on the brands. If it's Nestle, just normal products, I can talk about it [on social media]. But if it's too expensive and I always talk about it, it sounds like showing off."* A great deal of empathy for the issues of young people studying away from home was displayed by Dove chocolate in a very successful social media campaign, R6: *"Such as the Dove (chocolate brand) advertising when Chunyun* (the spring movement, when people from all over China are trying to go home and a large population is stuck in traffic*) starts, many students cannot get the tickets to go home. Dove created an activity to help students buying the ticket, in exchange student promote the Dove brand via @Dove account on Weibo, and then they randomly select few customers to buy tickets for them, so some students may not need to worry about how to buy a ticket to get home. I don't think the probability [I will win a ticket] is high, but I'll still forward it since it is a very easy to do it, by only clicking the button to win a chance or be the lucky person."* R12 summed it up *"I will be interested in things like...activities relating to my university life or about my study"*.

Overall, the proposition was supported. Frand relationships, and consumption in general are quite peripheral to the main issues and motivations of young urban students. Our respondents, unsurprisingly, were more concerned about their studies, getting home to see their families, and gaining acceptance and support from their peer group, then their latest purchases. Brands that showed empathy with the issues and preoccupations of the peer group – study, entertainment, youth culture, romance, and family ties were more likely to attract frands. There was also evidence that brands popular with youth e.g. snack foods, were more widely shared as people could be confident that their friends would also be interested, while more personal niche interests would be restricted to sharing within a smaller circle.

*P4: Effective frands will encourage personality expression and experimentation by both the brand and the customer.* Fun and interactive activities can also support self-expression. R6, who enjoyed Nike's competitions, commented: *"Maybe the judges don't like my design. But it doesn't influence my passion for participating. I have my own taste. I am not aiming to win the campaign. I like people to know my preference. Others will know what kind of shoes I like."*

Social media also offers a space where young Chinese consumers can experiment with global youth culture, R2: *"When I go to Starbucks with my friends, it seems that I belong to the new age instead of the old one...We always meet at Starbucks, so if they liked it [online], I will follow [Starbucks] too"*.

Less frequently studied is the extent to which social media allows brands to experiment with, and develop their identity. Chewing gum, for example, is a fairly commoditized product, not normally associated with a strong brand personality. Social media is used by a Chinese chewing gum brand to develop personality, R7: *"For example, Yida (Chinese gum), they shoot micro movies with four different themes (sour, sweet, bitter and spicy) as a series"*. This was one of only a few examples when a Chinese brand was discussed, and suggests that the effective social media strategies for foreign and Chinese brands have many similarities.

Social media can also assist with 'glocalization' and adding familiar symbols to less traditional products, R2: *"Nestle bought a large area in Yunnan province to produce their coffee beans, they posted the message saying 'we produce coffee in your country in Yunnan province',* *which makes me to think this international brand has the produce product in China, so I would like to pay attention to it and even try it since it makes me curious."*

Durex pursued a slightly edgy strategy that concentrated on local Chinese political figures and events, R4: *"Yes, I do follow some companies because the information the post are usually very interesting. Such as durex, they make fun of the recent most hot topics/drama such as political news in society that people are talking about to connect with their brand, which is very creative, and also advice about their own products."*

This overlaps, to some extent with the notion of compelling content, but picks up the theme of identity expression and experimentation. This proposition was largely supported, although experimentation seemed to be a stronger theme from brands than from customers. The separation of Chinese social media enables brands to experiment with developing 'glocal' personalities. Online competitions allowed customers to develop and express their taste. However, as we noted earlier in the context of peer groups, our respondents were very conscious of their online impression management, and unlikely to take risks with it by experimenting with a new frand. Exceptions included weakly connected and instrumental relationships, such as forwarding a post to gain a discount. Another example of experimentation was not so much at an individual level as a group level, where western brands such as Starbucks enabled young Chinese to experiment with a western coffee culture.

*P5: Effective frands will reflect and support users' motivations and priorities for using SNSs.* Our propositions are not mutually exclusive, so some of our respondent's views on this topic have already been captured. As we noted, brands tend to overestimate the degree to which customers seek brand intimacy and community membership on social media. Many followers are simply seeking interest (the IBM example) or using SNSs for entertainment (hence the popularity of micro-movies). Some are seeking information, R3: *"Sometimes…I just look for the price information."*

Brands can also assist and support users, without needing to be the centre of their attention. A keen runner liked Nike online, because Nike's posts made pursuing their social interest in running easier, R4: *"Because online media reaches lots of people. If I know there is a running event hold by Nike, I can just tell my friends through Facebook or Renren. I don't have to call them. So I can track lots of people. Maybe they will just run with me, because I post a message on Renren."*

People go on social media to communicate, be entertained, connect with friends, seek information, express themselves, pursue their interests, and (sometimes) to support their consumption behaviours. Brands that realise that they are not the main event, instead, the user, their social network, and their individual and collective motivations for using social media are the heart of the social media experience. Frands can be tolerated, even enjoyed and encouraged, to the extent that they recognize and support appropriate franding relationships. The more they behave like good friends, the more valued and better accepted they are likely to be.

## 4. Discussion and Conclusion

The role of brands in social media is extremely complex because people, and their use of social media are complex. We used a unifying metaphor of online friendship to explore online brand strategies and behaviours from foreign brands that attracted positive attention and engagement with young Chinese consumers.

Very few people would be naïve enough to imagine that they would be likely to make a lot of new, deep, genuine friends, using online channels only, in a foreign country. Yet the hype around social media marketing sometimes suggests this feat is not impossible for a brand. Our study confirms that social media, although it can be used for marketing, is not primarily a marketing space. Brands that respect this, and behave like good social media citizens – good frands will attract more awareness and positive engagement. Similarly, despite the amount of time many people spend on social media, it is only part of their life. Real life relationships, and life goals for work, study, leisure, friendship and romance profoundly influence social media behaviour. The vast majority of the online frand relationships engaged in by our respondents started off-line, and all the important ones did. Many of the brands that attracted comment about their social media presence had extensive off-line presence in China. Weak ties involving short-term, instrumental relationships are relatively easy to generate on social media (perhaps easier than using other channels) but it is unclear the extent to which these contribute towards long term loyalty, or endure beyond the life of the immediate offer.

Foreign brands have additional challenges, in deeply understanding the local context. Although many Western studies of social media have clear analogies in China, there are some important differences. However, foreign companies also have some advantages. Social media allows for some degree of experimentation in self-expression; a strategy a number of brands have used effectively.

Many companies hoping to enter the Chinese market are not large enough to have huge marketing budgets. The main contribution of our study is to not to advance a single theory, but to harmonize a wide variety of theoretical perspectives around a simple and intuitive metaphor that is effective in producing useful propositions. Although limited and exploratory at this stage, our propositions were well supported by our interview data. We suggest the following strategy for foreign brands developing a social media presence in China – don't expect too much, you are not your customer's best friend; but by being the best frand you can, both online and offline, you can gradually increase customer loyalty and engagement.